%Paper: hep-lat/9407021
%From: SASHA@UILGT1.PHYSICS.UIUC.EDU
%Date: Wed, 27 Jul 1994 16:22:27 -0500 (CDT)

%
%
%         figures 1 and 2 and the text are separated by ==============
%
%
\magnification=1200
\baselineskip=16pt
\rightline{ILL-(TH)-94-19}
\rightline{July, 1994}
\vskip1truecm

\centerline {\bf  CAN SIGMA MODELS DESCRIBE FINITE }
\centerline  {\bf  TEMPERATURE CHIRAL TRANSITIONS? }
\vskip1truecm

\centerline {Aleksandar KOCI\' C and John KOGUT}
\centerline {\it Loomis Laboratory of Physics, University of Illinois}
\centerline {\it 1110 W. Green St., Urbana, Il 61801-3080}
\vskip3truecm

\centerline {\bf Abstract}
\vskip 0.3truecm

Large-$N$ expansions and computer simulations indicate that the
universality class of the finite temperature chiral symmetry restoration
transition in the 3D Gross-Neveu model is mean field
theory. This is a counterexample to the standard 'sigma model'
scenario which predicts the 2D Ising model universality class.
We trace the breakdown of the standard scenario (dimensional
reduction and universality) to the absence of
canonical scalar fields in the model. We point out that our results could be
generic for theories with dynamical symmetry breaking, such as Quantum
Chromodynamics.

\vfill\eject

When studying the finite temperature chiral restoration transition in
QCD one is usually guided by the concepts of dimensional reduction
and universality.
A compelling idea, first put forward in [1] and later elaborated in
[2],
is that in four-dimensional $QCD$ with $N_f$ light quarks
%and, therefore,with $SU(N_f)_L \times SU(N_f)_R$ global symmetry group,
the physics near the chiral transition can be described by the
three-dimensional
$\sigma$-model with the same global symmetry.
The reasoning behind this proposal is based on counting the light degrees
of freedom and can be phrased as follows.
The transition region is dominated by the longitudinal
and transverse fluctuations of the order parameter, $\sigma$ and $\pi$,
which go soft at the transition temperature.
Being bosonic, $\sigma$ and $\pi$
have zero modes, $\omega_n =0$, in their finite-temperature
Matsubara decomposition. These zero modes
are the only relevant degrees of freedom in the scaling region and
at low energies the $n\not=0$ modes decouple.
Therefore, in the context of a $d$-dimensional theory,
one concludes that the phase transition
is described by an effective scalar theory in $d-1$ dimensions.
As a consequence, the chiral transition of
four-dimensional $QCD$, with $N_f=2$ flavors,
should lie in the same universality class as
a three-dimensional $O(4)$ magnet [1,2]. Similarly, other models
e.g. four-fermi theories in $d$-dimensions like
Gross-Neveu [3] with discrete or Nambu-Jona-Lasinio [4] with continuous
chiral symmetries, are expected to be in the universality class of a
$d-1$-dimensional Ising or Heisenberg magnet, respectively.

It is the purpose
of this paper to discuss the assumptions underlying this
analysis. As an illustration we will study
two examples: a purely bosonic theory, $O(N)$ $\sigma$-model where
the ideas of dimensional reduction apply, and
a Gross-Neveu model with composite scalars where they fail.
We discuss the generic features of the models that might apply to other
field theories at finite temperature.
At the end we comment on the implications these two examples have on $QCD$.

To illustrate how the idea of dimensional reduction is
realized in scalar theories,
we start with the $N$-component scalar theory and consider
the large-$N$ limit [5] for simplicity.
%$V(\phi )=1/2 \mu^2{\vec\phi}^2 +\lambda/4 ({\vec\phi}^2)^2$
To avoid
complications due to Goldstone bosons, we work in the symmetric phase.
At zero temperature, the susceptibility is given by the single tadpole
contribution.
%$$\chi^{-1}=\mu^2 +\lambda\int_q {1\over{q^2+\chi^{-1}} }\eqno(7)$$
Defining the critical curvature $\mu^2_c$ as the point where the
susceptibility diverges ($\mu_c^2+\lambda\int_q 1/q^2=0$), the
expression for the inverse susceptibility can be recast into

$$
\chi^{-1}\Biggl( 1+\lambda\int_q {1\over{q^2(q^2+\chi^{-1})} }\Biggr)
=\mu^2-\mu_c^2
\eqno(1)
$$
where $\int_q = \int d^dq/(2\pi )^d$, and
we absorb the combinatorial factor in $\lambda$.
The extraction of the critical index $\gamma$ reduces to counting powers
of the infrared (IR) singularities on the left hand side (LHS) of eq.(1).
Above four dimensions, both terms are IR finite and the
scaling is mean field ($\gamma =1$).
Below four dimensions the second term in eq.(1) dominates
the scaling region -- the integral diverges as $\chi^{(4-d)/2}$.
This gives the zero-temperature susceptibility exponent $\gamma=2/(d-2)$ [5].

At finite temperature, apart from the replacement of the frequency
integral with the Matsubara
sum, modifications are minimal [6]. For a given value of $\mu^2$
we define the critical temperature, $T_c$, by
$\mu^2 +\lambda T_c\sum_n\int_{\vec q} 1/(\omega_{nc}^2+{\vec q}^2)=0$,
%
%
%$$\chi^{-1}\Biggl( 1+\lambda T_c\sum_n\int_q
%{1\over{(\omega^2_{nc}+q^2)(\omega^2_n +q^2+\chi^{-1})} }\Biggr)
%$$
%$$
%=\lambda T_c\sum_n\int_q
%{{q^2(T/Tc-1)+\omega_n^2(T_c/T-1)}
%\over{(\omega^2_{nc}+q^2)(\omega^2_n +q^2+\chi^{-1})} }
%\eqno(9)
%$$
%
%
where $\omega_{nc}=2\pi nT_c$.
The momentum integrals are now performed over $d-1$ dimensional space.
Separating the $n=0$ mode $(\omega_0 =0)$
from the rest of the sum, we get the leading singular behavior

$$
\chi^{-1}\Biggl( 1+\lambda T_c \int_{\vec q}
{ 1\over{{\vec q}^2({\vec q}^2+\chi^{-1})} }+\sum_{n\not= 0}...
\Biggr)=
\lambda T_c\int_{\vec q}{  { T/T_c -1}\over{{\vec q}^2+\chi^{-1}} }
+\sum_{n\not= 0} ...
\eqno(2)
$$
The $n=0$ piece dominates the scaling region. It
resembles the zero-temperature expression,
eq.(1), except that now,
the integrals are performed in $d-1$ dimensions, instead of $d$. The power
counting is the same as before and it yields the thermal exponent
$\gamma_T=2/(d-3)$ which is the same as the zero-temperature $\gamma$ in
$d-1$ dimensions [6]. It is easy to obtain the other
critical exponents; they show the same type of behavior as $\gamma$.

To illustrate how compositeness affects the
physics near the phase transition, we analyze the problem of chiral symmetry
restoration in a Gross-Neveu model given by the lagrangian
$L=\bar\psi (i\partial +m +g\sigma )\psi -
{1\over 2}\sigma^2$,
where notation is standard [3].
Besides being an interesting theoretical model, it is also believed that,
when properly extended to incorporate continuous chiral symmetry, four-fermi
models are more realistic as effective theories of $QCD$ than the linear
sigma model, especially at scales where quark substructure is important.
When fermions are integrated out of the Gross-Neveu model, the Ising symmetry,
$\sigma\to -\sigma$, of the effective action  becomes manifest.
If the dimensional reduction + universality
arguments hold [1], the finite temperature transition of the $d$-dimensional
model would lie in the universality class of the $d-1$ dimensional
Ising model. In the remainder of the paper
we explain how and why this argument fails.

First, we start with the zero-temperature gap equation and
corresponding critical exponents.
The model can be treated in the
large-$N$ limit. To leading order, the fermion self-energy, $\Sigma$,
comes from the $\sigma$-tadpole: $\Sigma=m-g^2<\bar\psi\psi>$.
%$$\Sigma=m+4g^2\int_q {\Sigma\over{q^2+\Sigma^2}}\eqno(2)$$
To obtain the scaling properties of the theory, we define
the critical
coupling as $1=4g_c^2\int_q 1/q^2$. Combining this definition with the gap
equation leads to

$$
{m\over\Sigma}+\bigl( g^2/g_c^2 -1\bigr)=
4g^2\int_q {{\Sigma^2}\over{q^2(q^2+\Sigma^2)}}
\eqno(3)
$$
Like the scalar example,
this form is especially well suited for extracting critical indices
since the problem reduces again to the
counting of the infra-red divergences on the
right hand side [7,8]. The critical indices are defined by
$<\bar\psi\psi>|_{m=0}\sim t^\beta,
<\bar\psi\psi>|_{t=0}\sim m^{1/\delta},
\Sigma |_{m=0}\sim t^\nu,$ etc.. Here,
$t=g^2/g_c^2 -1$ is the deviation from the critical coupling.
Since $\Sigma\sim <\bar\psi\psi>$, $\beta=\nu$ to leading order.
Above four dimensions the integral in eq.(3) is finite
in the limit of vanishing $\Sigma$ and the scaling is mean-field.

Below four dimensions, the $\Sigma\to 0$
limit is singular -- the integral scales as
$\Sigma^{d-2}$.
Thus, in the chiral limit $t\sim \Sigma^{d-2}$, and at
the critical point, $t=0$,
away from the chiral limit, $m\sim\Sigma^{d-1}$.
The resulting exponents are non-gaussian: $\beta=1/(d-2)$ and
$\delta = d-1$. The remaining exponents are obtained
easily: $\eta=4-d, \gamma=1$ [7,8] and one can check that they
obey hyperscaling.

We now consider the Gross-Neveu model at finite-temperature.
We choose to stay between two and four dimensions
to emphasize how zero-temperature power-law scaling
changes at finite temperature. The gap equation is now modified to

$$
\Sigma=m+4Tg^2\sum_n\int_{\vec q} {\Sigma\over{\omega_n^2 +\vec q^2
+\Sigma^2}}
\eqno(4)
$$
where $\omega_n=(2n+1)\pi T$.
For $g>g_c$ the critical temperature is determined by:
$1=4T_c g^2\sum_n\int_{\vec q} 1/(\omega_{nc}^2+{\vec q}^2)$,
where $\omega_{nc}=(2n+1)\pi T_c$. This expression defines a critical
line in the $(g,T)$ plane.
For every coupling there exists
a critical temperature beyond which the symmetry is restored. Conversely,
for a fixed temperature there is a critical coupling, defined by the above
expression, corresponding to symmetry restoration.
At zero temperature, the symmetry is restored at $g=g_c$. Thus,
$(g=g_c, T=0)$ is the ultra-violet (UV) fixed point.
As the coupling moves away from $g_c$, a higher
restoration temperature results. At infinite coupling the end-point,
$(g=\infty , T=T_c)$, is the IR fixed point. The critical line
connects the UV and IR fixed points dividing the $(g,T)$ plane
into two parts.\footnote{$\,^{\dagger}$}{
The equation for the critical line can be brought into a
compact form by combining the expression for $T_c$ with the
definition of the zero-temperature critical coupling.  This results
in: $(g^2/g_c^2 -1) \sim T_c^{d-2}(g)$, i.e. $T_c(g) \sim \Sigma (T=0)$.
In this way, for any value of the coupling, the critical temperature
remains the same in physical units.}

Combining the definition of $T_c$ with the finite-temperature
gap equation, we can
bring it to a form similar to eq.(3)

$$
{m\over\Sigma}=\bigl( 1-T/T_c \bigr)+
4Tg^2\sum_n\int_{\vec q} {{\Sigma^2+\omega_{nc}(\omega_n+\omega_{nc})(T/T_c-1)}
\over{(\omega^2_{nc}+{\vec q}^2)(\omega_n^2+{\vec q}^2+\Sigma^2)}}
\eqno(5)
$$
The extraction of the critical exponents proceeds along
the same lines as in the zero-temperature case.  One difference
relative to eq.(3) becomes apparent immediately:
the zero modes are absent here and the integrand in
eq.(5) is regular in the $\Sigma\to 0$ limit even below four dimensions.
Consequently, the IR divergences are absent from all the integrals and
the scaling properties are those of mean-field theory:
 $\beta=\nu=1/2, \delta=3$, etc.
This is true for any $d$,  below or
above four. It appears that in this case, contrary to the scalar example,
the effect of making the
temporal direction finite ($1/T$)
is to regulate the IR behavior and
suppress fluctuations. This is
manifest in other thermodynamic quantities as well. For example,
to leading order, the scalar
susceptibility, $\chi =
\partial <\bar\psi\psi>/\partial m$, is given by

$$
\chi^{-1}=8g^2 T\sum_n\int_{\vec q} { {\Sigma^2}\over{(\omega_n^2+\vec
q^2+\Sigma^2)^2} }
\eqno(6)
$$
Once again, because of the absence of the zero mode ($\omega_0 =\pi T$),
the integral in eq.(6) is analytic in $\Sigma$, and the mean field relation
$\chi^{-1}\sim \Sigma^2$ follows. This is equaivalent to $\gamma=2\nu=1$.
The explicit calculation of the momentum dependence of the $\sigma$
propagator [9] yields $\eta=0$.

The scaling laws of the finite temperature transition
obtained above are completely different from
the predictions of ref.[1]. In fact, even the
systematics are opposite. The regulating character
of the temperature drives the lower dimensional theory towards
an effective theory that has gaussian critical exponents.

The fermionic model discussed above
was first analyzed in ref.[9]. Higher order calculations have shown
that the results are not artifacts of the large-$N$
limit [10]. In addition, it was explained in [10] how the Ising point is
recovered in four dimensional Yukawa models
beyond the leading order in $1/N$ and why this does not happen in
Gross Neveu models.
Lattice simulations of the three dimensional model have
verified the predictions of the large-$N$ expansion at zero temperature,
at nonzero temperature and at nonzero chemical potential [11].
The results for critical indices have been verified and improved
by larger scale simulations enhanced by histogram methods [12].
We have done additional simulations to check the finite temperature
results [11] in detail. Lattices of sizes $6\times30^2$,
$12\times36^2$ and $12\times72^2$ were simulated at $N = 12$
using the Hybrid Monte Carlo algorithm
described in [11]. High statistics runs (several tens of
thousands of trajectories for each coupling) were made on a variety of
lattices to guarantee that the simulations were probing the physical IR modes
at finite temperature. Luckily, our task is to distinguish mean field
exponents from those of the two dimensional Ising model, and, as reviewed in
Table I, they are dramatically different. We will discuss the exponents
$\delta$ and $\beta$ (defined above)
and leave other calculations to a lengthier
presentation. In Fig.1 we show the square of the order parameter $\sigma$
plotted against $1/g^2$. The data is in excellent
agreement with mean field theory where $\beta = 1/2$, and rules out the
Ising model value of $1/8$. Note that the statistical error bars in the
figure are smaller than the plotting symbols themselves. Since both lattice
sizes give the same estimates of $\beta$ while their critical
temperatures are quite different, we are confident that the simulation
is probing the true continuum behavior of the finite temperature
transition and is not corrupted by a sluggish crossover between symmetric
and asymmetric lattices. Several runs on the huge $12\times72^2$ lattice
were made and the $12\times36^2$ results were confirmed
to a fraction of a percent. We also calculated the susceptibility
index $\gamma$ and found $\gamma = 1.0(1)$ in the same runs. The Ising
result $\gamma = 7/4$ is decisively ruled out.
Next, we read off the
critical temperatures for both lattice sizes, measure the response of
the order parameter at criticality to an external symmetry breaking
field ( bare fermion mass ) and obtain $\delta$. The data is shown
in Fig.2, and for both lattice sizes we find $\delta = 3.1(1)$. The Ising
model value of $\delta = 15$ is ruled out. In all of these
calculations we carefully visualized the $\sigma$ field to check for
nonuniform configurations that would violate the mean field hypothesis [13].
None were found and all the past simulations [11] and the new ones reported
here support the contention that the large-$N$ results are reliable for this
problem.

An important feature of the
exponents corresponding to the finite-temperature transition,
is that they violate hyperscaling [14].
Usually, hyperscaling violation occurs above four dimension and is expressed
in terms of exponent inequalities [14] e.g.
$2\beta\delta - \gamma \leq d\nu$.
Strict inequality is applicable only
for $d>4$ and implies factorization of the correlation
functions. In our example the above
inequality goes in the opposite direction and
the breakdown of hyperscaling is not accompanied
by the factorization of Green's functions.

In conclusion, the study of the Gross-Neveu model
suggests that arguments invoking dimensional reduction
+ universality must be used with care.
Our results indicate that an effective scalar model fails to
describe the Gross-Neveu model at finite temperature.
We believe that the reason for this failure in our example is related
to the composite nature of the mesons.
Pointlike scalars can not adequately describe
the physics in the vicinity of the second order chiral transition.
The physical picture behind this failure observes that both the density
and the size of the loosely bound sigma meson increase with
temperature. Close to the restoration temperature
the system is densely populated with overlapping composites. In other
words the fluffiness of the mesons can not be ignored --
the constituent fermions are essential degrees of freedom even in the scaling
region, right before the composites dissociate.
Similar discussions of the failure of effective meson theories
in a slightly different context have been given in ref.[15].

It is well known that four-fermi models can be used as effective
theories of $QCD$ [16]. In addition to having the same global
symmetries, the mesons in both theories are composite. Therefore,
these models are believed to have common properties over a wide range of
scales where the quark substructure of the mesons is relevant.
Knowing this, it would be interesting to see what happens in two flavor
$QCD$ [17]: does it follow the dimensional reduction scenario,
or the Gross-Neveu behavior?
Of course, these two alternatives do not exhaust all the
possibilities [18], but we believe that the scenario suggested by the
Gross-Neveu model is sufficiently compelling to warrant further analyses
of $QCD$ simulation data.

We wish to acknowledge the discussions with Eduardo Fradkin and
Maria-Paola Lombardo. Special thanks go to Rob Pisarski for his comments on the
manuscript. This work is supported by NSF-PHY 92-00148
and used the computing facilities of PSC and NERSC.
\vfill\eject

\centerline{\bf References}

\noindent
[1] R. Pisarski and F. Wilczek, Phys. Rev. {\bf D29}, 338 (1984).

\noindent
[2] F. Wilczek, Int. J. Mod. Phys. {\bf A7}, 3911 (1992);
K. Rajagopal and F. Wilczek, Nucl. Phys. {\bf B404}, 577 (1993)

\noindent
[3] D. Gross and A. Neveu, Phys. Rev. {\bf D20}, 3235 (1974).

\noindent
[4] Y. Nambu and G. Jona-Lasinio, Phys. Rev. {\bf 122}, 345 (1961).

\noindent
[5] See, for example,
C.~Itzykson and J.-M.~Drouffe, {\it Statistical Field Theory} (Cambridge
University Press, 1989).

\noindent
[6] L. Dolan and R. Jackiw, Phys. Rev. {\bf D9}, 3320 (1974).

\noindent
[7] J. Zinn-Justin, Nucl. Phys. {\bf B367}, 105 (1991).

\noindent
[8]  S.~Hands, A.~Koci\'{c} and J.~B.~Kogut, Phys. Lett. {\bf B273}
(1991) 111.

\noindent
[9] B. Rosenstein, B. Warr and S. Park, Phys. Rev. {\bf D39}, 3088 (1989).

\noindent
[10] B. Rosenstein, A. Speliotopoulos and H. Yu,
Phys. Rev. {\bf D49}, 6822 (1994).

\noindent
[11] S.~Hands, A.~Koci\'{c} and J.~B.~Kogut, Ann. Phys. {\bf 224}, 29 (1993);
Nucl. Phys. {\bf B390} 355 (1993).

\noindent
[12] L. K\" arkk\" ainen, R. Lacaze, P. Lacock, and B. Petersson,
Nucl. Phys. {\bf B415}[FS], 781 (1994).

\noindent
[13] R. Dashen, S. K. Ma and R. Rajaraman, Phys. Rev. {\bf D11}, 1499 (1975);
F. Karsch, J. Kogut and H. W. Wyld, Nucl. Phys. {\bf B280}[FS18], 289 (1987).

\noindent
[14] B. Freedman and G.~A.~Baker Jr, J. Phys. {\bf A15} (1982) L715.
%R.~Schrader, Phys. Rev. {\bf B14} (1976) 172;
%B.~D.~Josephson, Proc. Phys. Soc. {\bf 92} (1967) 269, 276.

\noindent
[15] R. Jaffe and P. Mende, Nucl. Phys. {\bf B369}, 189 (1992).

\noindent
[16] See for example a recent review by T. Hatsuda nad T. Kunihiro,
Tsukuba preprint, UTHEP-270, (to appear in Phys. Rep.).

\noindent
[17] F. Karsch, Nucl. Phys. (Proc. Suppl.) {\bf B34}, 63 (1994).

\noindent
[18] E. Shuryak, Comments Nucl. Part. Phys. {\bf 21}, 235 (1994).

\vskip1truecm

\centerline{\bf Figure Captions}

\noindent
1. Order parameter squared plotted against temperature on
$6\times30^2$(left) and $12\times36^2$(right) lattices.

\noindent
2. Order parameter response at criticality plotted against bare fermion
mass on $6\times30^2$(bottom) and $12\times36^2$(top) lattices.

\vfill\eject

\baselineskip=20pt
\centerline{\bf Table 1}
\vskip 0.5 truecm
\centerline
{Critical exponents of the 3D Gross-Neveu and 2D Ising model}
\vskip 0.5 truecm
$$\vbox{\settabs\+\qquad18\qquad&\qquad190.4(8.8)\qquad&\qquad.1630(53)
\qquad&\cr

\+\hfill {}\hfill&\hfill $d=3$\hfill&\hfill Gross-Neveu
\hfill&\hfill $d=2$\hfill&\cr\smallskip
\+\hfill {}\hfill&\hfill $T=0$\hfill&\hfill$T\not= 0$\hfill&\hfill
Ising \hfill&\cr\smallskip
\+\hfill$\beta$\hfill&\hfill 1 \hfill&\hfill 1/2\hfill&\hfill
1/8\hfill&\cr\smallskip
\+\hfill$\delta$\hfill&\hfill 2\hfill&\hfill 3 \hfill&\hfill
15\hfill&\cr\smallskip
\+\hfill$\gamma$\hfill&\hfill1 \hfill&\hfill 1\hfill&\hfill 7/4
\hfill&\cr\smallskip
\+\hfill$\nu$\hfill&\hfill 1 \hfill&\hfill 1/2 \hfill&\hfill
1\hfill&\cr\smallskip
\+\hfill$\eta$\hfill&\hfill1 \hfill&\hfill 0 \hfill&\hfill 1/4
\hfill&\cr
\+\qquad$\,$\qquad&\qquad$\,$\qquad&\qquad$\,$\hfill&\cr}$$
%\vfill\eject
\vskip0.5truecm
\baselineskip=16pt

\end